\documentclass[a4paper,11pt]{article}

\usepackage{jheppub} 
\usepackage{etoolbox}                     
\makeatletter
\patchcmd{\maketitle}{\@fpheader}{}{}{}
\makeatother

\usepackage[T1]{fontenc}
\usepackage{amsmath,amssymb,mathtools,slashed}
\usepackage{color,xcolor}
\usepackage[utf8]{inputenc}
\usepackage{enumerate}
\usepackage[printonlyused]{acronym}
\usepackage{footnote}
\makesavenoteenv{tabular} 
\makesavenoteenv{table} 
\usepackage{hhline}
\usepackage{braket}
\usepackage{tikz}
\usepackage{tikz-feynman}
\usepackage[normalem]{ulem}

\usepackage{multirow}

 \definecolor{darkgreen}{rgb}{0.1,0.1,1.0}
 \hypersetup{
    linktocpage,
     colorlinks,
     citecolor=darkgreen,
     linkcolor= darkgreen,
     urlcolor=darkgreen
 }

\definecolor{blue}{rgb}{0,0,0.5}
\definecolor{dred}{rgb}{0.6,0.0,0.0}
\definecolor{dgreen}{rgb}{0.1, 0.5, 0.3}
\definecolor{dblue}{rgb}{0., 0.40, 0.8}

\definecolor{ivn}{rgb}{0.6,0.1,0.1}

\title{Nonperturbative Parameters of Inclusive $\Lambda_b$ Decays from Small Velocity Sum Rules}
 \author[a]{Bla\v zenka Meli\' c,}
 \author[a]{and Ivan  Ni\v sand\v zi\'c}

\affiliation[a]{Ru\dj er Bo\v skovi\'c Institute, Bijeni\v cka cesta 54, 10000, Zagreb, Croatia.}

\emailAdd{melic@irb.hr}
\emailAdd{ivan.nisandzic@irb.hr}

\abstract{
We investigate the nonperturbative parameters entering the heavy quark expansion for the inclusive decay rates of the $\Lambda_b$ baryon, focusing on the kinetic term $\hat{\mu}_\pi^2$ and the Darwin term $\hat{\rho}_D^3$. Recent evaluations have yielded an unexpectedly large Wilson coefficient associated with the dimension-six Darwin operator, which motivates a more detailed examination of the associated hadronic matrix element.
We constrain $\hat{\mu}_\pi^2$ and $\hat{\rho}_D^3$ using Small Velocity Sum Rules for the inclusive semileptonic decay $\Lambda_b \to X_c e^- \bar{\nu}_e$. These sum rules relate moments of hadronic structure functions, computed via the Operator Product Expansion, to hadronic matrix elements of relevant exclusive transitions. Truncating the hadronic side to include the lowest-lying charmed baryon states with spin-parity $ J^P = 1/2^+, 1/2^-, 3/2^- $, we employ the corresponding lattice QCD form factors near zero recoil as inputs. By analyzing zeroth and higher moments, we derive inequalities that define an allowed region in the $(\hat{\mu}_\pi^2, \hat{\rho}_D^3)$ plane. The derived constraints are consistent, within uncertainties, with estimates from spectroscopic relations and the nonrelativistic constituent quark model. This work refines the determination of key HQE parameters for $\Lambda_b$, with implications for precision predictions of heavy baryon lifetimes and inclusive decays of heavy baryons in general.}

\date{\today}

\begin{document}
\preprint{RBI-ThPhys-2025-24}
\maketitle
\flushbottom

\section{Introduction} 
Heavy quark expansion (HQE) provides a well-developed framework for study of weak inclusive decays of heavy hadrons. It involves a systematic expansion of the inclusive rates in powers of the inverse of the heavy quark mass. In a number of recent advances, this expansion has been extended to higher orders including $\alpha_s$ corrections \cite{Mannel:2019qel,Lenz:2020oce,Mannel:2020fts,Moreno:2020rmk,Fael:2020tow,Piscopo:2021ogu,Czakon:2021ybq,Mannel:2021zzr,Mannel:2023zei,Moreno:2024bgq,Mannel:2025fvj}. Recent investigations have also revisited quark-hadron duality violations \cite{Mannel:2024crj}, and the treatment of the charm quark mass \cite{Boushmelev:2023kmf}.

Alongside these improvements, accurate predictions require knowledge of nonperturbative forward matrix elements of the effective operators generated by the expansion. These matrix elements have been extracted by fitting the theoretical predictions to available experimental data, see e.g.~\cite{Alberti:2014yda,Bordone:2021oof,Bernlochner:2022ucr,Finauri:2023kte} for recent analyses, and \cite{Bernlochner:2024vhg} for future prospects. Other methods include QCD sum rules, and the use of spectroscopic mass relations which arise from applying the HQE to heavy hadron masses. In principle, the relevant matrix elements could also be evaluated using nonperturbative lattice techniques.

A series of recent papers have presented up-to-date predictions for the lifetimes of heavy hadrons, including heavy mesons \cite{King:2021xqp, Gratrex:2022xpm, Lenz:2022rbq}, and heavy baryons \cite{Gratrex:2022xpm, Gratrex:2023pfn, Dulibic:2023jeu}. These analyses have incorporated the contributions of the recently determined Wilson coefficient for the dimension-six Darwin operator \cite{Mannel:2020fts, Lenz:2020oce, Moreno:2020rmk, Piscopo:2021ogu}. Interestingly, this coefficient appears significantly larger than naively expected—much larger than the corresponding Wilson coefficients of dimension-five contributions. This motivates a more thorough examination of the corresponding hadronic matrix element. Existing estimates typically rely on a relation, derived from the equation of motion for the gluon field strength, to the matrix elements of four-quark operators~\cite{Pirjol:1998ur, Czarnecki:1997sz}. In the case of heavy mesons, the latter matrix elements can be estimated using the vacuum-insertion approximation or the sum-rules techniques \cite{Kirk:2017juj,King:2021jsq,Black:2024bus}. Meanwhile, the Darwin parameter for the $B$ meson has also been extracted using the fits to the experimental data for the inclusive semileptonic decays~\cite{Alberti:2014yda,Bordone:2021oof,Bernlochner:2022ucr,Finauri:2023kte}. 
For heavy baryons, however, the experimental data is less complete. Inclusive semileptonic decay widths have been measured only for $\Lambda_c$ baryon, no differential spectra are available for either charm or bottom baryons, and for $b$ baryons the inclusive semileptonic widths remain unmeasured, although their lifetimes are known.
In this case, the Darwin parameter is estimated via four-quark matrix elements that have been obtained using the nonrelativistic constituent-quark model (NRCQM) applied to spectroscopic data for heavy baryons \cite{Blok:1991st, Neubert:1996we,Rosner:1996fy,Guberina:1997yx, Gratrex:2022xpm,Gratrex:2023pfn}.

In this paper, we study the constraints on the kinetic and Darwin parameters for the $\Lambda_b$ baryon using available lattice QCD results for form factors describing the exclusive transitions ($\Lambda_b \to \Lambda_c^{1/2^+}$, $\Lambda_c^{1/2^-}$, $\Lambda_c^{3/2^-}$) near zero recoil~\cite{Detmold:2015aaa, Meinel:2021mdj, Meinel:2021rbm}. Specifically, we aim to determine the region in the space of these parameters consistent with the constraints coming from the small velocity sum rules (SVSRs) in the semileptonic inclusive decays of the $\Lambda_b$ baryon.  Matrix elements of remaining chromomagnetic and spin-orbit dimension-five operators vanish for the $\Lambda_b$ due to the spin structure of its constituents.

This paper is organised as follows. In Section~\ref{Background}, we review the essential theoretical background, covering the relevant parts of HQE formalism for inclusive semileptonic decays and defining the nonperturbative parameters $\hat{\mu}_\pi^2$ and $\hat{\rho}_D^3$. Section~\ref{Small velocity sum rules} details our main analysis using SVSRs. We outline the well-known derivation of various SVSRs, the usage of lattice QCD inputs for the exclusive $\Lambda_b$ transition form factors, and the construction of specific combinations of moments of the hadronic structure tensor. In Section \ref{Numerical Analysis of the Constraints}, we describe our numerical procedure for constraining the HQE parameters. We compare our results with existing estimates and provide short conclusions in Section~\ref{Discussion and conclusions}. Finally, Appendices~\ref{Hadronic tensor coefficients} and \ref{App: Form factor parametrizations} provide key technical details on the OPE coefficients and form factor parametrisations, respectively.

\section{Background}\label{Background}
We consider the semileptonic decay process $\Lambda_b \to X_c e \bar{\nu}_e$ where the final state $X_c$ is inclusive over appropriate singly charmed hadronic states. The weak effective Hamiltonian for the underlying $b\to c e\bar{\nu}$ transitions is
\begin{equation}
    \mathcal{H}^{cb}_{\text{SL}} = \frac{G_F}{\sqrt{2}} V_{cb}\,J^\alpha\,(\bar{\ell} \gamma_\alpha (1-\gamma_5) \nu_\ell) + \text{h.c.},
\end{equation}
with the implicit sum over charged lepton flavours $\ell$, and with the quark current
\begin{equation}
    J_\alpha(x) = \bar{c}(x) \gamma_\alpha (1 - \gamma_5) b(x)\,.\label{Eq:DefofCurrent}
\end{equation} 

The hadronic part of the contributions to the inclusive semileptonic decay rate is encoded in the structure tensor
\begin{equation}
    W_{\alpha\beta} \equiv (2\pi)^4\int_{X_c}\delta^{(4)}(p_{\Lambda_b}-p_{X_c}-q)\frac{\bra{\Lambda_b}J_\alpha^\dagger (0)\ket{X_c}\bra{X_c}J_\beta (0)\ket{\Lambda_b}}{2m_{\Lambda_b}}\,,\label{Eq:DefW}
\end{equation}
where the symbol $\int_{X_c}$ denotes an inclusive sum over all hadronic final states $X_c$ containing a single charm quark. This includes discrete sums over one-particle and multi-particle states, sums over polarizations, and the corresponding Lorentz-invariant phase-space integrals over the final-states momenta\footnote{For example, for one-particle states $\ket{X_i}$, we have:
\begin{equation*}
    \int_{X_i|\text{1-particle}} = \sum_{i,s} \int \frac{d^3 \mathbf{p}_{X_i}}{(2\pi)^3 2E_{X_i}}\,.
\end{equation*}}. 
The structure tensor is conventionally parametrised as
\begin{equation}
W_{\alpha\beta}=-w_1 g_{\alpha\beta}+w_2 v_\alpha v_\beta
-iw_3 \epsilon_{\alpha\beta\rho\sigma}v^\rho q^\sigma
+w_4 q_\alpha q_\beta+w_5(q_\alpha v_\beta+q_\beta v_\alpha)\,,\label{Eq:WParametrization}
\end{equation}
in terms of invariant coefficients $w_i = w_i(q^2, q\cdot v)$, where $v=p_{\Lambda_b}/m_{\Lambda_b}$ denotes the four-velocity of the $\Lambda_b$ baryon and $q$ is the four-momentum transferred to the lepton pair. Subsequently, we work in the rest-frame of $\Lambda_b$ baryon where the kinematic variables are expressed in terms of $q_0$ and $\mathbf{q}$, with $\mathbf{p}_{X_c}=-\mathbf{q}$, where $\mathbf{p}_{X_c}$ denotes the spatial momentum of the final hadronic state.

The hadronic structure tensor is related to the imaginary part of the forward matrix element of the tensor $h_{\alpha\beta}$ which is defined as the matrix element of the time-ordered product of two insertions of the quark currents:
\begin{equation}
h_{\alpha\beta} = i \int d^4x \, e^{-i q \cdot x} \frac{\bra{\Lambda_b} T\{J_\alpha^\dagger(x) J_\beta(0)\} \ket{\Lambda_b}}{2m_{\Lambda_b}} \,. \label{Eq:hab}
\end{equation}
This tensor can be calculated using the operator product expansion (OPE) in the kinematic region away from its discontinuities.
In analogy with the structure tensor, $h_{\alpha\beta} $ is parametrised in terms of coefficients $h_i = h_i(q^2, q \cdot v)$ as
\begin{equation}
h_{\alpha\beta} = -h_1 g_{\alpha\beta} + h_2 v_\alpha v_\beta - i h_3 \epsilon_{\alpha\beta\rho\sigma} v^\rho q^\sigma + h_4 q_\alpha q_\beta + h_5 (q_\alpha v_\beta + q_\beta v_\alpha)\,. \label{Eq:hParametrization}
\end{equation}

In the following, we will separately consider the spatial and temporal contractions of the weak currents in the rest frame of the $\Lambda_b$ baryon~\cite{Bigi:1994ga}. For this purpose, we use the projectors~\cite{Leibovich:1997az}:
\begin{equation}
s^{\alpha\beta} = -g^{\alpha\beta} + v^\alpha v^\beta \,, \qquad \qquad t^{\alpha\beta} = v^\alpha v^\beta \,,\label{Eq:Projectors}
\end{equation}
such that, in the rest frame of the $\Lambda_b$ baryon, $s^{\alpha\beta} h_{\alpha\beta}$ projects to the spatial contraction
\begin{equation}
s^{\alpha\beta} h_{\alpha\beta} = h_{kk} = 3 h_1 + h_4\,\mathbf{q}^2\,, \label{Eq:ProjectorsSpatial}
\end{equation}
with the implicit sum over the repeated spatial index $k$, while $t^{\alpha\beta}$ projects to the time-time component:
\begin{equation}
t^{\alpha\beta} h_{\alpha\beta} = h_{00} = -h_1 + h_2 + h_4\,q_0^2 + 2h_5\,q_0\,.
\end{equation}

By inserting a complete set of states between the two currents in the time-ordered product in $h_{\alpha\beta}$, one finds that this tensor exhibits branch cuts along the real axis in the complex $q_0$ plane for a fixed $\vec{q}$. One such cut, which corresponds to the contribution with $\theta(x^0)$, occurs for $q_0=m_{\Lambda_b}-E_{X_c}$, where $E_{X_c}=(m_{X_c}^2+\vert\vec{q}\vert^2)^{1/2}$ denotes the energy of the state $X_c$. It includes the semileptonic decay region with contributions from charmed states produced in the semileptonic decay of $\Lambda_b$ and extends from $q_{0\,\text{min}}=\vert\vec{q}\vert$ to $q_{0\,\text{max}}=m_{\Lambda_b}-E_{\Lambda_c}$, where the latter limit corresponds to the production of the lowest-lying charmed baryon $\Lambda_c$. Another region, adjacent to the semileptonic cut, extends from $-\infty$ to $0$. A separate, distant cut, which corresponds to the contribution with $\theta(-x^0)$ begins at $
(m_{X_{\bar{c}bb,\text{min}}}^2+|\mathbf{q}|^2)^{1/2}-m_{\Lambda_b}$ and extends to infinity. It receives contributions from multiparticle states with the flavour content $\bar{c}bbud$ \cite{Manohar:2000dt}. 

\begin{figure}[h]
\centering
\begin{tikzpicture}[xscale=2.2, yscale=2.8]
    \draw (-1.6, 0) -- (2.4, 0) node[right] {Re$(\epsilon)$};  
    \draw (0, -1.) -- (0, 1.) node[above] {Im$(\epsilon)$}; 

    \draw[ultra thick, dgreen, dotted] (-1, 0.035) -- (-1.6, 0.035);

    \draw[ultra thick, dblue] (0, -0.035) -- (1.4, -0.035);

    \draw[ultra thick, dred, dashed] (1.4, -0.035) -- (2.4, -0.035);

    \node at (-0.37, -0.025) {\small *}; 
    \node at (-0.37, -0.1) {$\epsilon_0$}; 

    \draw[thick, black, postaction={decorate}, decoration={markings, mark=at position 0.56 with {\arrow{latex}}}] 
    (2.4, -0.135) -- (0, -0.135);

    \draw[thick, black] (0,-0.135) arc[start angle=270, end angle=90, radius=0.1];

    \draw[thick, black, postaction={decorate}, decoration={markings, mark=at position 0.5 with {\arrow{latex}}}] 
    (0, 0.065) -- (2.4, 0.065);
    
    \node[black] at (1.2, -0.28) {$\mathcal{C}_1$};
\end{tikzpicture}
\caption{Singularity structure of $h_{\alpha\beta}$ for a fixed value of $\mathbf{q}$ in the complex $\epsilon$-plane: Semileptonic decay cut (solid blue), adjacent cut (dashed red), distant cut (dotted green). Contour enveloping the cut on the positive real axis is denoted by $\mathcal{C}_1$. The semileptonic cut receives contributions from charmed states $X_c$ in $\Lambda_b\to X_c\ell\bar{\nu}$ and extends from $\epsilon_{\text{min}}=0$ to $\epsilon_{\text{max}}=m_{\Lambda_b}-E_{\Lambda_c}-\vert\vec{q}\vert$. The point $\epsilon_0$ (marked by $*$) indicates a location on the negative real axis, sufficiently far from the cuts and with $|\epsilon_0| \gg \Lambda_{\text{QCD}}$ where the OPE can be applied.}
\label{Fig:CutsEps1Contour}
\end{figure}
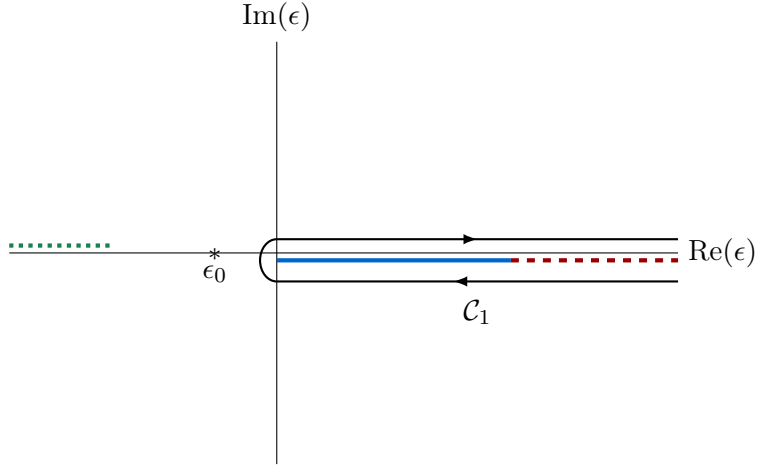

It is then customary to introduce the variable $\epsilon$, defined as the difference between $m_{\Lambda_b} - E_{\Lambda_c}$ which represents the endpoint of the semileptonic cut, and $q_0$:
\begin{equation}
    \epsilon \equiv (m_{\Lambda_b} - E_{\Lambda_c}) - q_0\,.
    \label{Eq:Epsil}
\end{equation}
On the semileptonic cut, $\epsilon$ corresponds to the energy of a state $X_c$ above the lowest-lying charmed hadron, that is, $\epsilon = E_{X_c} - E_{\Lambda_c}$. The branch cuts of the tensor $h_{\alpha\beta}$ the complex-$\epsilon$ plane are illustrated in Fig.~\ref{Fig:CutsEps1Contour}.

Using Cauchy's integral formula, one can express $h_i(\epsilon_0)$ at a point $\epsilon_0$ away from the singularities as a dispersion integral over a contour circumventing the cuts. A convenient choice is to place $\epsilon_0$ on the negative real axis, with $|\epsilon_0| \gg \Lambda_{\text{QCD}}$ sufficiently far from the distant cut so that the contribution of this cut can be neglected. Under this assumption, the dispersive integral reduces to an integral over the contour $\mathcal{C}_1$, see Fig.~\ref{Fig:CutsEps1Contour}. It is assumed that the tensor at the point $\epsilon_0$ is given by the corresponding OPE expression, $h_i(\epsilon_0)=h_i^{(\text{OPE})}(\epsilon_0)$. Restricting the integral along the positive real axis to $\epsilon < \Delta$, with $\Lambda_{\text{QCD}} \ll \Delta$, and expanding the dispersion integral in powers of $1/\epsilon_0$, one obtains 
\begin{equation}
\frac{1}{\pi} \int_0^\Delta d\epsilon\, \epsilon^n\, \text{Im}\, h_i^{(\text{OPE})}(\epsilon) = \frac{1}{2\pi} \int_0^\Delta d\epsilon\, \epsilon^n\, w_i(\epsilon)\,,\label{Eq:DispOPE}
\end{equation}
where $h_{\alpha\beta}^{(\text{OPE})}$ represents the analytic continuation of the OPE result to the semileptonic cut. On the right-hand side $w_{i}=1/i\,\operatorname{Disc}\,h_{i}=2\operatorname{Im}h_{i}$
appears as the discontinuity of the 
tensor $h_{i}$ across the cut. The OPE is therefore expected to capture the moments of the hadronic structure tensor, encoding averaged information about the nonperturbative contributions of the hadronic states.

To systematically organize the moments of the hadronic structure tensor, we use the following notation. We denote the $n$-th moment of a given combination of coefficients in the structure tensor as
\begin{equation}
\mathcal{I}_a^{\Gamma_1 \Gamma_2\,(n)}\equiv \frac{1}{2\pi}\int_0^\Delta\,d\epsilon\,\epsilon^n a^{\alpha\beta}W_{\alpha\beta}^{\Gamma_1 \Gamma_2}\,,\label{Eq: MomentDefi}
\end{equation}
where $\Gamma_{1}$ and $\Gamma_2$ are the Dirac structures of the inserted currents (e.g., vector or axial-vector), and $a^{\alpha\beta}$ is projector $a^{\alpha\beta}=s^{\alpha\beta}$ or $a^{\alpha\beta}=t^{\alpha\beta}$, defined in Eq.~\eqref{Eq:Projectors}. For example, the $n$-th moment that includes the spatial components of the axial-vector currents we have
\begin{equation}
    \mathcal{I}_{kk}^{AA\,(n)} \equiv \frac{1}{2\pi}\int_0^\Delta 
    d\epsilon\,\epsilon^n s^{\alpha\beta}\,W_{\alpha\beta}^{AA}\,,
\end{equation}  
with the implicit summation over the repeated spatial index $k$.

The OPE expressions for the relevant coefficients $h_i$, evaluated up to $\mathcal{O}(\Lambda_{\text{QCD}}^3)$ and leading order in $\alpha_s$ \cite{Gremm:1996df,Koyrakh:1996fr,Dassinger:2006md,Colangelo:2020vhu}, are listed in equations~\eqref{Eq:app1}–\eqref{Eq:app6} of Appendix~\ref{Hadronic tensor coefficients}.
In these expressions, $q_0$ is to be written in terms of the hadronic excitation energy $\epsilon$ defined in Eq.~\eqref{Eq:Epsil}, as
\begin{equation}
    q_0 = \delta + m_b - E_c - \epsilon\,,
\end{equation}
where $E_c = (m_c^2+|\vec{q}|^2)^{1/2}$, and
\begin{equation}
    \delta = (m_{\Lambda_b} - m_b) - (E_{\Lambda_c} - E_c)\,.
\end{equation}
The term $\delta$ can then be expanded in powers of $1/E_c$ and $1/m_b$, see e.g.~\cite{Bigi:1994ga}.
For the zeroth, second, and third moments of structure functions used in this work, contributions from  expansion of $\delta$ enter at $\mathcal{O}(\Lambda_{\text{QCD}}^4)$ and are thus neglected, consistently with the $\mathcal{O}(\Lambda_{\text{QCD}}^3)$ to which we work.

In the OPE expansion up to $\mathcal{O}(\Lambda_{\text{QCD}}^3)$  we will encounter matrix elements of nonvanishing two-quark operators for $\Lambda_b$, up to dimension six, which we define here, following Refs.~\cite{Dassinger:2006md,Gratrex:2022xpm, Gratrex:2023pfn}.
We have, for a general b-hadron\footnote{We assume the standard relativistic normalization of hadronic states,
$\braket{H_b(p')|H_b(p)} = (2\pi)^3 2E_{H_b}\,\delta^{(3)}(\vec{p} - \vec{p}\,')$.}:
\begin{align}
\hat{\mu}_\pi^2(H_b) &= -\frac{1}{2M_{H_b}} \langle H_b(p) | \bar{b}_v (i D_\mu) (i D^\mu) b_v | H_b(p) \rangle \,,
\label{eq:mu-PI} \\
\hat{\mu}_G^2(H_b) &= \frac{1}{2M_{H_b}} \langle H_b(p) | \bar{b}_v (i D_\mu) (i D_\nu) (-i \sigma^{\mu \nu}) b_v | H_b(p) \rangle \,,
\label{eq:mu-G} \\
\hat{\rho}_D^3(H_b) &= \frac{1}{2M_{H_b}} \langle H_b(p) | \bar{b}_v (i D_\mu) (i v \cdot D) (i D^\mu) b_v | H_b(p) \rangle \,, \label{eq:rho-D}
\end{align}
with the phase-redefined b-quark field $b_v(x) \equiv \exp(i m_b v \cdot x) b(x)$. The parameters are denoted with a hat to make a distinction to analogous quantities defined in Heavy Quark Effective Theory (HQET), where parameters without hats are defined through matrix elements involving the HQET field $h_v$. To order $\mathcal{O}(\Lambda_{\text{QCD}}^3/m_b^3)$, we have ${\hat \mu}_\pi^2 = \mu_\pi^2$ and ${\hat \rho}_D^3 = \rho_D^3$. Concerning ${\hat \mu}_G^2$, the relation between the two definitions is given~\cite{Colangelo:2020vhu, Dassinger:2006md}, to $\mathcal{O}(1/m_b)$, as:
\begin{equation}
    {\hat \mu}_G^2 = \mu_G^2 - \frac{1}{m_b} \left( \rho_D^3 + \rho_{\text{LS}}^3 \right) \,,
\end{equation}
where the spin-orbit matrix element $\rho^3_{\text{LS}}$ is defined as
\begin{equation}
    \hat{\rho}^3_{\text{LS}} \equiv \frac{1}{2M_{H_b}} \langle H_b(p) | \bar{b}_v (i D_\mu) (i v \cdot D) (i D_\nu) (-i \sigma^{\mu\nu}) b_v | H_b(p) \rangle \,.
\end{equation}

Furthermore, we recall that $\mu_G^2(H_b)\propto d_H$, where $d_H$ is the appropriate spin factor, see e.g. \cite{Neubert:1993mb} for the definitions. This factor is vanishing for $\Lambda_b$ baryon, so that 
\begin{equation}
    \mu_G^2(\Lambda_b)=0\,,
\end{equation}
and, similarly, $\rho_{\text{LS}}^3(\Lambda_b)=0$. Therefore, our analysis will focus on constraining the two relevant non-perturbative HQE parameters: the kinetic term $\hat{\mu}^2_\pi$ and the Darwin term $\hat{\rho}^3_D$, using inputs from lattice QCD and experimental data.

\section{Small velocity sum rules}\label{Small velocity sum rules}
The SVSRs for the semileptonic decays of heavy hadrons can be traced back to the early 1990s~\cite{Bjorken:1990hs,Isgur:1990jf,Isgur:1991wr,Voloshin:1992wg,Bigi:1994re,Bigi:1994ga,Chow:1994ac,Chow:1995mz,Bigi:1997fj,Uraltsev:1998bk,Uraltsev:1996rd,Uraltsev:2000fi}, with Ref.~\cite{Bigi:1994ga} providing an early systematic account. A detailed discussion of their application for inclusive $B$ meson decays can be found in~\cite{Gambino:2012rd}, see also \cite{Boyd:1996hy,Davoudiasl:1997yf}. Applications to semileptonic $\Lambda_b$ baryon decays were explored in Ref.~\cite{Leibovich:1997az} and, more recently, in Ref.~\cite{Mannel:2015osa}.

To relate the OPE expressions to the contributions of specific hadronic states, we now turn to the hadronic representation of the structure tensor. Equation~\eqref{Eq:DispOPE} relates the moments of the structure tensor obtained from the OPE, to the corresponding moments obtained by including a finite number of lowest-lying hadronic states in the expression for $w_i$ on the right-hand side.
Near the zero-recoil point, which corresponds to small $|\mathbf{q}|$, these relations yield the SVSRs. Truncating the hadronic sum to a few lowest-lying states turns the SVSRs into inequalities. These inequalities can then be used either to constrain the form factors for the exclusive semileptonic transitions near zero recoil or, conversely, to provide the constraints on HQE matrix elements when independent information on the exclusive form factors is available. In our analysis, we adopt the second approach: we use the lattice QCD determinations of the form factors for the exclusive semileptonic transitions $\Lambda_b\to\Lambda_c^{1/2^+},\quad \Lambda_b\to\Lambda_c^{1/2^-},\quad \text{and} \quad \Lambda_b\to\Lambda_c^{3/2^-}$ near zero recoil~\cite{Detmold:2015aaa, Meinel:2021mdj, Meinel:2021rbm} as inputs to constrain the HQE parameters $\hat{\mu}^2_\pi$ and $\rho^3_D$.

With the explicit inclusion of the lowest-lying charmed baryons $X_c =\{ \Lambda_c^{1/2^+}, \Lambda_c^{1/2^-},\Lambda_c^{3/2^-} \}$ in the structure tensor \eqref{Eq:DefW}, we have:
\begin{equation}
W_{\alpha\beta}^{\Gamma_1\Gamma_2} = (2\pi)^4 \sum_{s,s',X_c} \int \frac{d^3p_{X_c}}{(2\pi)^3 2E_{X_c}}\, \delta^{(4)}(p_{\Lambda_b} - q - p_{X_c})\, \frac{\langle \Lambda_b(s) \vert J_{\alpha}^{\dagger\,\Gamma_1} \vert X_c(s') \rangle \langle X_c (s') \vert J_\beta^{\Gamma_2} \vert \Lambda_b(s) \rangle}{4m_{\Lambda_b}} + \ldots\,,
\label{Eq:TruncatedW}
\end{equation}
with the additional factor of $1/2$ which accounts for the sum over the spin projections 
$s$ of $\Lambda_b$. Strictly speaking, among the three lowest-lying states, only the ground state baryon $\Lambda_c^{1/2+}$ is a one-particle (bound) state in QCD. Nevertheless, it is reasonable to treat the narrow resonances $\Lambda_c^{1/2^-}$ and $\Lambda_c^{3/2^-}$ as effective one-particle states capturing the contributions from the multiparticle final states into which they decay. The ellipsis in Eq.~\eqref{Eq:TruncatedW} represents contributions from higher excitations and from the multiparticle continuum which remain largely unconstrained. The spectrum of higher charm resonances, with masses lighter than about $3$ GeV, has been estimated in the relativistic quark-diquark picture in Ref.~\cite{Ebert:2007nw} and is listed in Table~\ref{Tab:1} below.

Inserting $q_0 = m_{\Lambda_b} - E_{\Lambda_c} - \epsilon$ and performing the integration over the spatial momenta, we have
\begin{equation}
    \mathcal{I}_a^{\Gamma_1\Gamma_2\,(n)} = \int_0^\Delta d\epsilon\,\epsilon^n\, \sum_{X_c}\delta\big(\epsilon-(E_{X_c}-E_{\Lambda_c})\big) \frac{\mathcal{F}_{a,X_c}^{\Gamma_1\Gamma_2}}{8 m_{\Lambda_b} E_{X_c}} + \ldots\,.
\end{equation}
where
\begin{equation}
\mathcal{F}_{a,X_c}^{\Gamma_1\Gamma_2}\equiv \sum_{s,s'} a^{\alpha\beta}\langle\Lambda_b(s)\vert J_{\alpha}^{\dagger\,\Gamma_1}\vert X_c(s')\rangle \langle X_c(s')\vert J_\beta^{\Gamma_2}\vert \Lambda_b(s)\rangle\,.
\end{equation}
Finally, performing the integral over $\epsilon$ and using \eqref{Eq:DispOPE} leads to the sum rule
\begin{equation}
\begin{split}
     \frac{1}{2\pi}\int_0^\Delta d\epsilon\,\epsilon^n\, a^{\alpha\beta}W_{\alpha\beta}^{\Gamma_1\Gamma_2\,\text{(OPE)}}
     &= \sum_{X_c}\,(E_{X_c}-E_{\Lambda_c})^n \frac{\mathcal{F}_{a,X_c}^{\Gamma_1\Gamma_2}}{8 m_{\Lambda_b} E_{X_c}} + \ldots\,
\end{split}
\label{Eq:SumRule}
\end{equation}
where the factor on the right-hand side includes implicitly the step-function prefactor $\theta(\Delta-(E_{X_c}-E_{\Lambda_c}))\theta(E_{X_c}-E_{\Lambda_c})$. Since each term on the right-hand side of the sum rule is positive, restricting the exclusive sum to the three lowest-lying states converts the equalities into corresponding inequalities such that the OPE expressions for the moments provide {\it upper bounds} on the corresponding included hadronic contributions.

\subsection{Zeroth moments}
At zero recoil point ($\vec{q} = 0$) the zeroth moments computed via the operator product expansion (OPE) up to $\mathcal{O}(\Lambda_{\text{QCD}}^3)$ take the following form. For the vector current, we have:
\begin{equation}
\mathcal{I}_{kk}^{VV\,(0),\,\text{OPE}}\big\vert_{\text{0-recoil}} = \frac{\hat{\mu}_\pi^2}{4}\bigg(\frac{3}{m_c^2}+\frac{3}{m_b^2}-\frac{2}{m_b m_c}\bigg)
-\frac{\hat{\rho}_D^3}{4}\bigg(\frac{3}{m_c^3}-\frac{3}{m_b^3}-\frac{1}{m_b m_c^2}+\frac{1}{m_b^2 m_c}\bigg)\,,\label{Eq:OPEkkVV}
\end{equation}
and
\begin{equation}
\mathcal{I}_{00}^{VV\,(0),\,\text{OPE}}\big\vert_{\text{0-recoil}} = \xi_V
-\frac{\hat{\mu}_\pi^2}{4}\left(\frac{1}{m_c}-\frac{1}{m_b}\right)^2
-\frac{\hat{\rho}_D^3}{4}\left(\frac{1}{m_c}-\frac{1}{m_b}\right)^2\left(\frac{1}{m_c}+\frac{1}{m_b}\right)\,,\label{Eq:OPE00VV}
\end{equation}
for spatial and time components, respectively, while for the axial current
\begin{equation}
\begin{split}
\mathcal{I}_{kk}^{AA\,(0),\,\text{OPE}}\big\vert_{\text{0-recoil}} &= 3\Bigg[\xi_A
-\frac{\hat{\mu}_\pi^2}{4}\left(\frac{1}{m_c^2}+\frac{1}{m_b^2}+\frac{2}{3m_b m_c}\right)\\
&-\frac{\hat{\rho}_D^3}{4}\left(\frac{1}{m_c^3}+\frac{1}{3m_b m_c^2}+\frac{1}{m_b^3}+\frac{1}{3m_b^2 m_c}\right) \Bigg]\,,
\end{split} \label{Eq:OPEkkAA}
\end{equation}
and
\begin{equation}
\mathcal{I}_{00}^{AA\,(0),\,\text{OPE}}\big\vert_{\text{0-recoil}} = \frac{\hat{\mu}_\pi^2}{4}\left(\frac{1}{m_b}+\frac{1}{m_c}\right)^2
-\frac{\hat{\rho}_D^3}{4}\left(\frac{1}{m_c^2}-\frac{1}{m_b^2}\right)\left(\frac{1}{m_c}+\frac{1}{m_b}\right)\,.\label{Eq:OPE00AA}
\end{equation}
Here, the coefficients $\xi_{V}$ and $\xi_{A}$ include $\alpha_s$ corrections for the vector~\cite{Uraltsev:2003ye} and axial-vector~\cite{Gambino:2010bp} coefficients. The sum rule for the moment $\mathcal{I}_{kk}^{VV}$ has been previously considered to $\mathcal{O}(\Lambda_{QCD}^2$) in the context of constraining the kinetic parameter in Ref.~\cite{Leibovich:1997az}.

On the other hand, the form factors for the exclusive semileptonic transition $\Lambda_b \to \Lambda_c$ have been computed using lattice QCD in Ref.~\cite{Detmold:2015aaa}. We use the results from the analysis of Ref.~\cite{Meinel:2021rbm}, which incorporates the zero-recoil consistency relations among the form factors as pointed out in Ref.~\cite{Hiller:2021zth}. The form factors for transitions to the negative-parity excited states $\Lambda_b \to (\Lambda_c^{1/2^-}, \Lambda_c^{3/2^-})$ have been evaluated in Refs.~\cite{Meinel:2021mdj, Meinel:2021rbm} to first order in $(w - 1)$, and are parametrized schematically as:
\begin{equation}
    f_i = F^{f_i} + (w - 1) A^{f_i}\,,\label{Eq:ParaemtrisW}
\end{equation}
where $w \equiv v \cdot v'$, and $(w - 1) \simeq \vec{q}^{\,2}/(2M^2)$, with $M$ denoting the mass of a final-state charmed baryon under consideration. The $w$-dependence away from the zero recoil can be systematically described using parametrizations derived from HQET~\cite{Leibovich:1997az,Korner:2000zn}, with recent explorations including the lattice form factors given in Refs.~\cite{Boer:2018vpx,Papucci:2021pmj}. In the present analysis, we do not rely on an HQET expansion for the exclusive form factors but instead use the expansion around zero recoil schematically represented by Eq.~\eqref{Eq:ParaemtrisW}, adopting the explicit results from the lattice QCD calculations. The corresponding parametrisations of the exclusive matrix elements in terms of form factors are collected in Appendix~\ref{App: Form factor parametrizations} (Eqs.~\eqref{FF1}--\eqref{FF10}).
The nonvanishing contributions from the three lowest-lying states to the zeroth hadronic moments at zero recoil are given, for the vector-vector current contractions as
\begin{align}
    \mathcal{I}_{kk}^{VV\,(0),\,\text{had}}\big|_{\text{0-recoil}} &= 3\left| f_{+}^{(1/2^-)}(w=1) \right|^2 + 2\left| f_{\perp'}^{(3/2^-)}(w=1) \right|^2\,, \label{Eq:Ikkhad1} \\
    \mathcal{I}_{00}^{VV\,(0),\,\text{had}}\big|_{\text{0-recoil}} &= \left| f_{0}^{(1/2^+)}(w=1) \right|^2\,, \label{Eq:Ikkhad2}
\end{align}
while for the axial-vector contractions:
\begin{align}
    \mathcal{I}_{kk}^{AA\,(0),\,\text{had}}\big|_{\text{0-recoil}} &= 3\left| g_{+}^{(1/2^+)}(w=1) \right|^2\,, \label{Eq:Ikkhad3} \\
    \mathcal{I}_{00}^{AA\,(0),\,\text{had}}\big|_{\text{0-recoil}} &= \left| g_{0}^{(1/2^-)}(w=1) \right|^2\,. \label{Eq:Ikkhad4}
\end{align}
The corresponding zeroth-moment sum rules take the form of inequalities:
\begin{equation}
    \mathcal{I}_{kk}^{\Gamma\Gamma\,(0),\,\text{OPE}}\big\vert_{\text{0-recoil}} > \mathcal{I}_{kk}^{\Gamma\Gamma\,(0),\,\text{had}}\big\vert_{\text{0-recoil}}\,, \qquad \qquad
    \mathcal{I}_{00}^{\Gamma\Gamma\,(0),\,\text{OPE}}\big\vert_{\text{0-recoil}} > \mathcal{I}_{00}^{\Gamma\Gamma\,(0),\,\text{had}}\big\vert_{\text{0-recoil}}\,,
    \label{Eq:ZerothInequalities}
\end{equation}
with $\Gamma=A, V$.

\subsection{Higher Moments}
Further constraints on the hadronic matrix elements can be obtained using higher moments of the structure tensor. In particular, we consider the second moment $\mathcal{I}_{kk}^{AA(2)}$ and the third moment $\mathcal{I}_{kk}^{AA(3)}$, involving the spatial contractions of the axial currents. The linear terms in the corresponding $|\vec{q}|^2$-expansions around zero recoil provide, at leading order in $\alpha_s$ and $\Lambda_{\text{QCD}}$, direct access to the kinetic and the Darwin terms.

The corresponding OPE expressions, expanded to first order in $|\vec{q}|^2$ and to $\mathcal{O}(\Lambda_{\text{QCD}}^3)$ are given as
\begin{equation}
\mathcal{I}_{kk}^{AA(2)\,,\text{OPE}} =\frac{|\vec{q}|^2}{2m_c^2}\left[ 2\hat{\mu}_\pi^2+\frac{8\alpha_s}{3\pi}\,\Delta^2- \hat{\rho}_D^3\left(\frac{1}{m_c} + \frac{1}{3m_b}\right)\right]\,,
\end{equation}
and
\begin{equation}
\mathcal{I}_{kk}^{AA(3),\,\text{OPE}}=\frac{|\vec{q}|^2}{2m_c^2}\bigg(2\hat{\rho}_D^3+\frac{16\alpha_s}{9\pi}\,\Delta^3\bigg)\,,
\end{equation}
including $\alpha_s$ corrections \cite{Davoudiasl:1997yf} to the leading order in $1/m_Q$. We further employ the analogous expression \cite{Davoudiasl:1997yf} for the zeroth moment $\mathcal{I}_{kk}^{AA(0)}$, given as
\begin{align}
\mathcal{I}_{kk}^{AA(0),\,\text{OPE}} =\; \mathcal{I}_{kk}^{A A\,(0),\,\text{OPE}}\big\vert_{\text{0-recoil}}+\frac{|\vec{q}|^2}{2m_c^2}&\Bigg[
\; -\frac{3}{2}
-\frac{8\alpha_s(\mu_s)}{9\pi}\Big(5 - 3\log\frac{4\Delta^2}{\mu_s^2}\Big)\notag\\
&+\hat{\mu}_\pi^2\Big(\frac{15}{4 m_c^2}+\frac{3}{4m_b^2} + \frac{5}{6m_b m_c}\Big)\notag\\
& 
+\hat{\rho}_D^3\Bigl(\frac{11}{4m_c^3}+\frac{3}{4 m_b m_c^2}+\frac{5}{12m_b^2 m_c}+\frac{3}{4m_b^3} \Bigr)
\Biggr]\,,
\end{align}
where the leading term on the right-hand side is given in Eq. \eqref{Eq:OPEkkAA}.
We recall once again that $\Delta$ denotes the upper limit of the integration in the moments (see Eq.~\eqref{Eq:DispOPE}). For the scale of strong coupling $\alpha_s$, we use $\mu_s = \sqrt{m_b m_c}$, following the prescription in Ref.~\cite{Uraltsev:1994sc}.

As for the hadronic side, we note that the contributions from the two lowest-lying parity-odd states to $\mathcal{I}_{kk}^{AA\,(2)\,,\text{had}}$ and $\mathcal{I}_{kk}^{AA\,(3)\,,\text{had}}$ are proportional to $\vec{q}^{\,2}$ and thus vanish at the zero-recoil point. Consequently, the derivatives of these moments over $\vert \vec{q}\vert^2$ depend solely on the values of the form factors at the zero recoil, which is a welcome feature since the coefficients of the leading-order expansion of the lattice QCD form factors at zero recoil exhibit smaller uncertainties compared to those at the linear order.

Rather than considering the second and third moments individually, we construct suitable combinations of the moments to cancel the contributions of specific excited states, to reduce uncertainties arising from that excited state. A similar approach has been used for inclusive $B$ meson decays (see e.g. Refs.~\cite{Davoudiasl:1997yf,Boyd:1996hy}), where combinations of the zeroth and the second moments were used to remove the effect of the first excited state. We introduce the following combinations
\begin{equation}
    Y_n \equiv \frac{\mathcal{I}_{kk}^{AA\,(2)}}{(E_n-E_0)^2} - \mathcal{I}_{kk}^{AA(0)}\,,\qquad \qquad Z_n \equiv \frac{\mathcal{I}_{kk}^{AA (3)}}{E_n-E_0} - \mathcal{I}_{kk}^{AA(2)}\,,
    \label{eq:YZmoments}
\end{equation}
where $n$ denotes the subtracted state, with $n=0$ corresponding to the ground state baryon $\Lambda_c$, while $n = 1,2$ denote the first and second excited states $\Lambda_c^{1/2^-}$ and $\Lambda_c^{3/2^-}$, respectively. In particular, we employ $Y_2$, which involves subtracting the contribution of the second excited state $\Lambda_c^{3/2^-}$. Using Eq.~\eqref{Eq:SumRule}, we have
\begin{equation}
\begin{split}
    Y_2& = \frac{\mathcal{I}_{kk}^{AA(2)}}{(E_{\Lambda_c^{3/2^-}}-E_{\Lambda_c})^2} - \mathcal{I}_{kk}^{AA(0)}\\
    &=-\frac{\mathcal{F}_{kk,\Lambda_c}^{AA}}{8m_{\Lambda_b} E_{\Lambda_c}}+\Bigg(\frac{(E_{\Lambda_c^{1/2^-}}-E_{\Lambda_c})^2}{(E_{\Lambda_c^{3/2^-}}-E_{\Lambda_c})^2}-1\Bigg)\frac{\mathcal{F}_{kk,\Lambda_c^{1/2^-}}^{AA}}{8m_{\Lambda_b} E_{\Lambda_c^{1/2^-}}}+\ldots \label{Eq:Y3}
\end{split}
\end{equation}
where the ellipsis denotes the contributions from higher excited states starting with the third excitation.
The second and third moments vanish at zero recoil, so we use linear terms in the expansion around zero recoil. The positivity of the derivatives of the neglected terms over $|\vec{q}|^2$ at zero recoil \cite{Davoudiasl:1997yf} leads to the inequality:
\begin{equation}
   \frac{dY_2^{\text{OPE}}}{d\vert\vec{q}\vert^2}\Bigg\vert_{\vert\vec{q}\vert^2=0}>\frac{d}{d\vert\vec{q}\vert^2}\Bigg[-\frac{\mathcal{F}_{kk,\Lambda_c}^{AA}}{8m_{\Lambda_b} E_{\Lambda_c}}+\Bigg(\frac{(E_{\Lambda_c^{1/2^-}}-E_{\Lambda_c})^2}{(E_{\Lambda_c^{3/2^-}}-E_{\Lambda_c})^2}-1\Bigg)\frac{\mathcal{F}_{kk,\Lambda_c^{1/2^-}}^{AA}}{8m_{\Lambda_b} E_{\Lambda_c^{1/2^-}}}\Bigg]\Bigg\vert_{\vert\vec{q}\vert^2=0}\,,
   \label{Eq:SRY3}
\end{equation}
where the hadronic inputs on the right-hand side are available from lattice QCD results near zero recoil~\cite{Detmold:2015aaa, Meinel:2021mdj, Meinel:2021rbm}. 

The analogous combination of the second and third moments is
\begin{equation}
\begin{split}
    Z_2& = \frac{\mathcal{I}_{kk}^{AA(3)}}{E_{\Lambda_c^{3/2^-}}-E_{\Lambda_c}} - \mathcal{I}_{kk}^{AA(2)}\\
    &=(E_{\Lambda_c^{1/2^-}}-E_{\Lambda_c})^2\bigg(\frac{E_{\Lambda_c^{1/2^-}}-E_{\Lambda_c}}{E_{\Lambda_c^{3/2^-}}-E_{\Lambda_c}}-1\bigg)\frac{\mathcal{F}_{kk,\Lambda_c^{1/2^-}}^{AA}}{8m_{\Lambda_b}E_{\Lambda_c^{1/2^-}}}+\ldots\,,
\end{split}
\end{equation}
with the remaining truncated term starting with the contribution from the third excited state. Since all other contributions from higher excited states to the higher moments vanish at zero recoil and are positive for $|\vec{q}|^2 > 0$, their derivatives over $|\vec{q}|^2$ at zero recoil are also positive, which leads to another inequality
\begin{equation}
   \frac{dZ_2^{\text{OPE}}}{d\vert\vec{q}\vert^2}\Bigg\vert_{\vert\vec{q}\vert^2=0}>\frac{d}{d\vert\vec{q}\vert^2}\Bigg[(E_{\Lambda_c^{1/2^-}}-E_{\Lambda_c})^2\bigg(\frac{E_{\Lambda_c^{1/2^-}}-E_{\Lambda_c}}{E_{\Lambda_c^{3/2^-}}-E_{\Lambda_c}}-1\bigg)\frac{\mathcal{F}_{kk,\Lambda_c^{1/2^-}}^{AA}}{8m_{\Lambda_b}E_{\Lambda_c^{1/2^-}}}\Bigg]\Bigg\vert_{\vert\vec{q}\vert^2=0}\,. \label{Eq:SRZ3}
\end{equation}

\begin{figure}[ht]
    \centering
\includegraphics[width=0.5\textwidth]{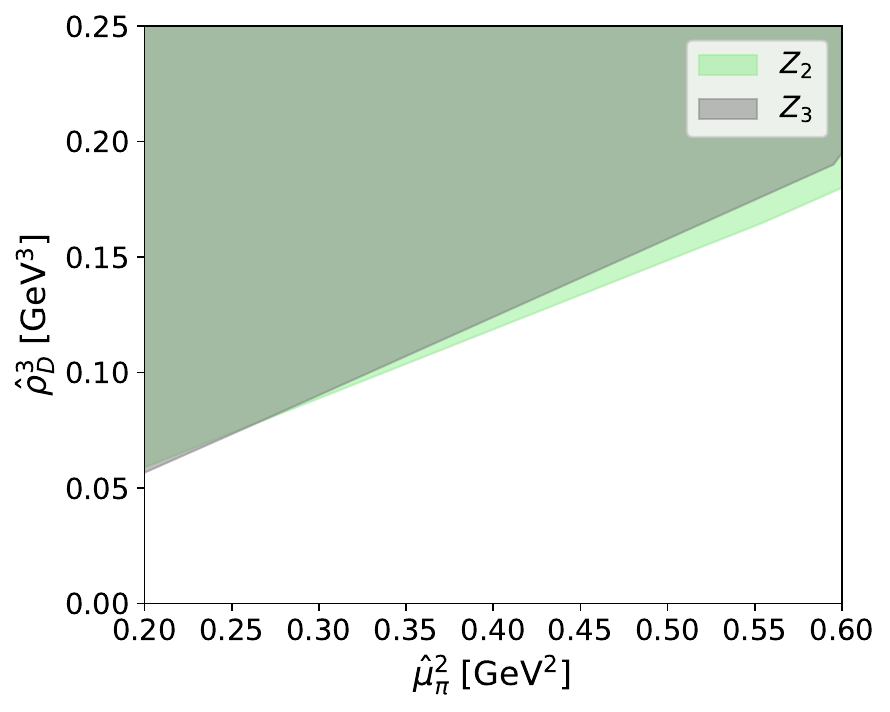}
    \caption{Region in the $(\hat{\mu}_\pi^2, \hat{\rho}_D^3)$ plane that satisfies the sum-rule inequalities for  $\frac{dZ_2}{d\vert\vec{q}\vert^2}$ (light green) and $\frac{dZ_3}{d\vert\vec{q}\vert^2}$ (gray).}
    \label{Fig:PlotBigZ2Z3}
\end{figure}

For our final analysis, we will use inequalities involving $Y_2$ and $Z_2$. Although this choice omits a state for which lattice results are available, it allows us to formulate constraints using a minimal set of nonperturbative inputs.
We have also considered the corresponding inequalities obtained using $Y_3$ and $Z_3$, as defined in Eq.~\eqref{eq:YZmoments}, and compared the resulting constraints with those derived from $Y_2$ and $Z_2$. The combinations $Y_3$ and $Z_3$ subtract the contribution of the third excited state, which has not yet been experimentally confirmed. The relativistic quark–diquark model~\cite{Ebert:2007nw} provides an estimate for its mass, $M(\Lambda_c^{1/2^+(2S)})=2.77\,\text{GeV}$, see Table~\ref{Tab:1}. A candidate resonance with the mass $2766.6\,\text{MeV}$ has been observed~\cite{CLEO:2000mbh}, though its quantum numbers remain unknown, and it may correspond to the $\Sigma_c$ state instead.
To account for this uncertainty, we varied the mass of the third excited state within a $75\,\text{MeV}$ range around the model's estimate. 
In Fig.~\ref{Fig:PlotBigZ2Z3} we compare a simultaneous constraint on $\hat{\mu}_\pi^2$ and $\hat{\rho}_D^3$, stemming from the $Z_2$ and $Z_3$-inequalities 
and demonstrate the consistency between these two. Similarly, it can be shown that there is no significant improvement of the constraint when inequalities involving $Y_2$ and $Y_3$ are compared. These findings give us confidence that we can use, in our final analysis only $Y_2$ and $Z_2$ to obtain reliable results.

\begin{table}[h!]
\centering
\caption{The mass spectrum of the isospin singlet charmed baryons, denoted by $M$, lighter than about $3$ GeV, according to the relativistic quark-diquark picture in Ref.~\cite{Ebert:2007nw}. Comparison is made with the available confirmed experimental values $M_{\text{exp}}$ \cite{ParticleDataGroup:2024cfk}.}
\vspace{0.3cm}
\begin{tabular}{|c|c|c|c|}
\hline
$J^P$ & State & $M$ (MeV) & $M_{\text{exp}}$ (MeV) \\
\hline
$1/2^+$ & $1S$ & $2297$ & $2286.46(14)$ \\
$1/2^-$ & $1P$ & $2598$ & $2592.25(28)$ \\
$3/2^-$ & $1P$ & $2628$ & $2628.00(15)$ \\
$1/2^+$ & $2S$ & $2772$ &  \\
$3/2^+$ & $1D$ & $2874$ & $2856.1^{+2.3}_{-6.0}$ \\
$5/2^+$ & $1D$ & $2883$ &  $2881.63(24)$\\
$1/2^-$ & $2P$ & $3017$ &  \\
\hline
\end{tabular}
\label{Tab:1}
\end{table}

\section{Numerical analysis of the constraints}\label{Numerical Analysis of the Constraints}
In this section, we detail our numerical procedure for constraining the hadronic parameters $\hat{\mu}_\pi^2$ and $\hat{\rho}_D^3$ by comparing the values of the moments evaluated in the OPE with the corresponding hadronic quantities evaluated using the lattice QCD results for the form factors of the lowest-lying charm baryons.

For our analysis we adopt the kinetic mass scheme \cite{Bigi:1994ga, Bigi:1996si}, where heavy quark pole masses are recast as short-distance, $\mu$-dependent kinetic masses, with $\mu$ ($\sim 1\,\text{GeV}$) acting as a Wilsonian cutoff separating soft and hard gluon contributions. Specifically, this amounts to replacing the pole mass in our expressions with the kinetic mass using the one-loop relation:
\begin{equation}
m_{c, \rm pole} = m_{c,\mathrm{kin}}(\mu) \left[ 1 + \frac{4\alpha_s}{3\pi}\left(\frac{\mu}{m_{c,\mathrm{kin}}(\mu)} + \frac{1}{2}\frac{\mu^2}{m_{c,\mathrm{kin}}(\mu)^2}\right) \right]\,,
\end{equation}
and expressing the HQE parameters in terms of $\mu$-dependent quantities defined in the kinetic scheme, with the corresponding perturbative parts subtracted~\cite{Bigi:1994ga, Bigi:1996si}, as $\hat{\mu}_\pi^2 = \hat{\mu}_\pi^2(\mu) - \left[\hat{\mu}_\pi^2(\mu)\right]_{\mathrm{pert}}$ and $\hat{\rho}_D^3 = \hat{\rho}_D^3(\mu) - \left[\hat{\rho}_D^3(\mu)\right]_{\mathrm{pert}}$, see e.g. Ref.~\cite{Fael:2024rys} for a detailed discussion. The radiative corrections can be found in \cite{Czarnecki:1997sz, Fael:2020njb}; in the current analysis, we use the corresponding expressions to order $\alpha_s$. Furthermore, in accord with the kinetic scheme, we identify the cutoff of the integration $\Delta$ in our expressions with the scale $\mu$. In what follows, and unless otherwise specified, all HQE parameters are quoted in the kinetic scheme with cut-off of $\mu = 0.75~\mathrm{GeV}$ \cite{Mannel:2015osa}.

The kinetic masses have been evaluated in terms of corresponding values in the $\overline{\text{MS}}$ scheme up to $\mathcal{O}(\alpha_s^3)$ in Ref.~\cite{Fael:2020njb} with the results implemented in the Wolfram Mathematica package \verb+RunDec+ \cite{Chetyrkin:2000yt,Herren:2017osy}. To maintain consistency with our calculation's order of precision in $\alpha_s$, we use the one-loop conversion from the $\overline{\text{MS}}$ scheme using the $\overline{\text{MS}}$ values $\overline{m}_c(\overline{m}_c)=1.28\,\mathrm{GeV}$ and $\overline{m}_b(\overline{m}_b)=4.18\,\mathrm{GeV}$ as inputs, to obtain:
\begin{equation}
    m_{c,\,\rm kin}(0.75\,\text{GeV})=1.29\,\text{GeV}\,,\qquad \qquad  m_{b,\,\rm kin}(0.75\,\text{GeV})=4.48\,\text{GeV}.
\end{equation}

We apply the inequalities ~\eqref{Eq:SRY3} and \eqref{Eq:SRZ3} for the derivatives of $Y_2$ and $Z_2$ over $|\vec{q}|^2$ at the zero recoil, as well as the inequalities~\eqref{Eq:ZerothInequalities} for the zeroth moments at the zero recoil. The corresponding hadronic quantities, evaluated using the lattice data, are collected as:
\begin{equation}
\mathbf{M}^{\mathrm{had}} \equiv \Biggl( \left.\frac{dY_2}{d|\vec{q}|^2}\right|_{|\vec{q}|^2=0},\; \left.\frac{dZ_2^{\mathrm{had}}}{d|\vec{q}|^2}\right|_{|\vec{q}|^2=0},\; \mathcal{I}_{kk}^{VV\,(0),\mathrm{had}},\; \mathcal{I}_{kk}^{AA\,(0),\mathrm{had}},\; \mathcal{I}_{00}^{VV\,(0),\mathrm{had}},\; \mathcal{I}_{00}^{AA\,(0),\mathrm{had}} \Biggr)\,.
\end{equation}
The corresponding array of the OPE expressions as functions of $\hat{\mu}_\pi^2$ and $\hat{\rho}_D^3$ is denoted by $\mathbf{V}(\hat{\mu}_\pi^2, \hat{\rho}_D^3)$.
Since the hadronic quantities obtained from the lattice calculations are subject to statistical uncertainties, we treat them as random variables distributed with probability distribution function $p(\mathbf{M}^{\mathrm{had}},\mathcal{C}^{\text{had}})$ where $\mathcal{C}^{\text{had}}$ denotes the corresponding covariance matrix. We estimate this distribution by evaluating $\textbf{M}^{\text{had}}$ from a large number of samples of the form factors based on the covariance matrices for the form-factor coefficients provided in the auxiliary files of the lattice evaluations in Refs.~\cite{Detmold:2015aaa, Meinel:2021mdj}, assuming these coefficients follow the corresponding multi-variable Gaussian distribution. For each pair of values of $(\hat{\mu}_\pi^2, \hat{\rho}_D^3)$, we evaluate the joint probability that a randomly drawn array of hadronic moments, distributed with the distribution $p(\mathbf{M}^{\mathrm{had}}, \mathcal{C}^{\mathrm{had}})$, lies below the corresponding OPE prediction $\mathbf{V}(\hat{\mu}_\pi^2, \hat{\rho}_D^3)$. This probability is given by the cumulative distribution function
\begin{equation}
\mathrm{CDF}\big(\mathbf{V}(\hat{\mu}_\pi^2, \hat{\rho}_D^3)\big) 
= \int_{-\infty}^{\mathbf{V}(\hat{\mu}_\pi^2, \hat{\rho}_D^3)} p(\mathbf{M}^{\text{had}})\, d^6\mathbf{M}^{\text{had}}\,,
\label{Eq:CDFdef}
\end{equation}
where the upper limit is understood component-wise. We then define the allowed region in the $(\hat{\mu}_\pi^2, \hat{\rho}_D^3)$ plane as the set of parameter points for which this function exceeds a chosen threshold $P_{\mathrm{th}}$.

\begin{figure}[ht]
    \centering
\includegraphics[width=0.51\textwidth]{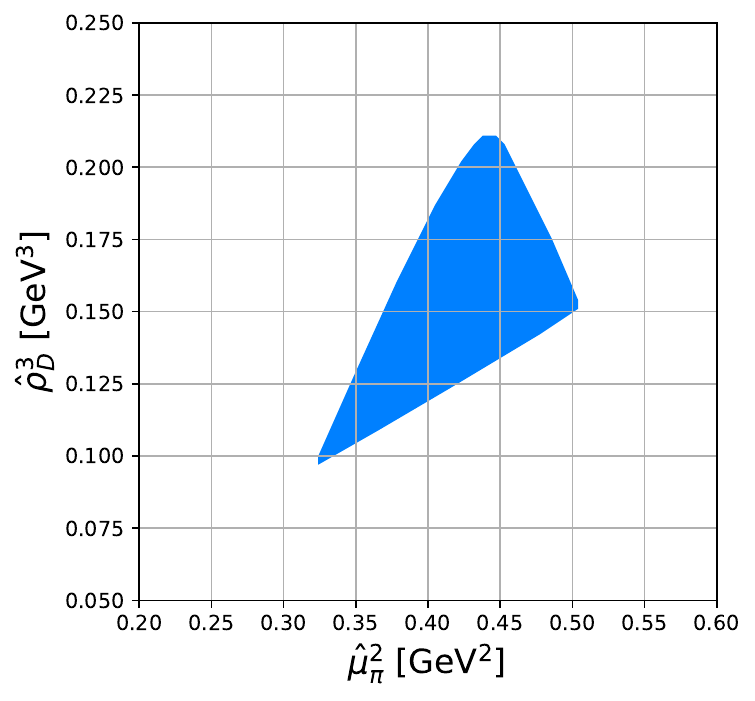}
    \caption{Allowed region in the $(\hat{\mu}_\pi^2, \hat{\rho}_D^3)$ plane where the sum-rule inequalities are satisfied. This region is defined by imposing that the cumulative probability—derived from matching lattice QCD determinations of the hadronic moments with the corresponding OPE predictions—exceeds a $50\%$ threshold (see the text in Sec.~\ref{Numerical Analysis of the Constraints} for more details). All values are given in the kinetic scheme with the
      cut-off scale $\mu = 0.75~\text{GeV}$.}
    \label{Fig:PlotBig}
\end{figure}

\begin{figure}[ht]
    \centering
\includegraphics[width=0.51\textwidth]{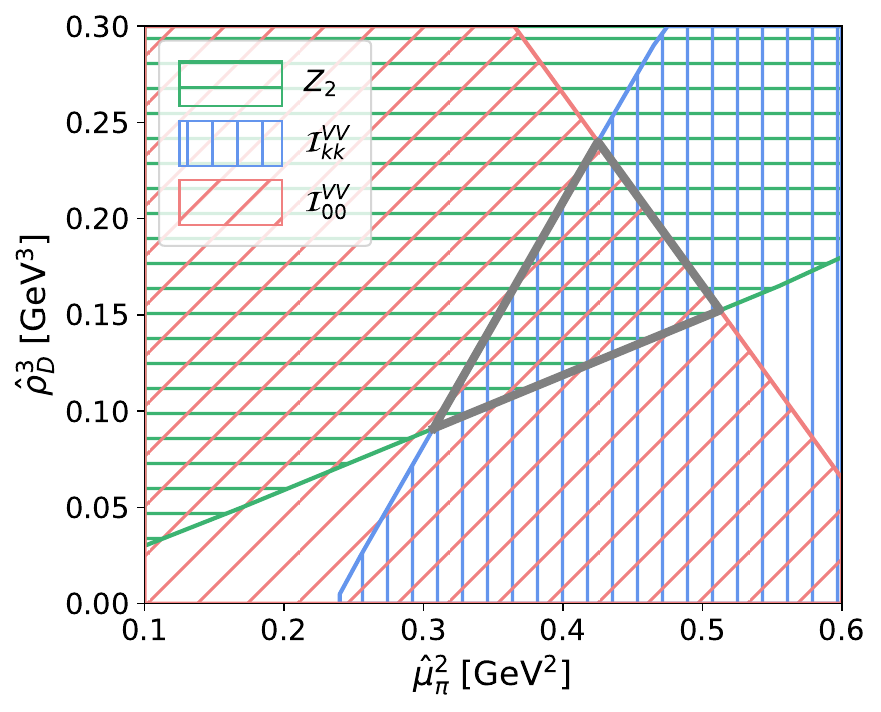}
   \caption{Interplay of the three most constraining inequalities. The allowed region in the $(\hat{\mu}_\pi^2,\hat{\rho}_D^3)$ plane is defined by requiring the OPE predictions to exceed the central lattice values of the hadronic moments. $Z_2$ denotes the derivative of the relevant moment at zero recoil (see Eq.~\eqref{Eq:SRZ3}).}
    \label{Fig:Combination}
\end{figure}

We find that, for all the values of $\hat{\mu}_\pi^2$ and $\hat{\rho}_D^3$ that are consistent with the remaining constraints, the OPE values for $\mathcal{I}_{00}^{VV}$ and $\mathcal{I}_{kk}^{AA}$ do not significantly exceed the central values of the corresponding lattice QCD results. The latter moments turn out to be nearly saturated by the ground-state $\Lambda_c$ contribution in the corresponding parameter region. This observation is consistent with findings in Ref.~\cite{Mannel:2015osa}, where the authors observed the saturation using the value of the kinetic term estimated from heavy-hadron spectroscopy and the value of the Darwin term that has been adopted from the fits for inclusive $B$ meson decays.
We therefore adopt a probability threshold of $P_{\text{th}} = 50\%$ when defining the allowed region in the $(\hat{\mu}_\pi^2,\, \hat{\rho}_D^3)$ plane.
This value is dictated by $\mathcal{I}_{00}^{VV}$ since within the aforementioned parameter space, the OPE prediction for this moment barely exceeds the corresponding hadronic central value.
Then the corresponding allowed region is shown in Fig.~\ref{Fig:PlotBig}. Since there exists the kinetic and Darwin parameter region for which the $\mathcal{I}_{00}^{VV}$ moment lies within the one-sigma band of the corresponding lattice result, the observed near saturation does not presently imply an inconsistency, given the unknown magnitude of contributions from higher excited states. However, it introduces a strong sensitivity to the lattice values of this moment, so that the improved precision in the lattice determination of $\mathcal{I}_{00}^{VV}$ could be beneficial. One may also achieve a comprehensive understanding of the contributions from states higher than $\Lambda_c$. 

Following the statistical determination of the allowed region (Fig.~\ref{Fig:PlotBig}), in Fig.~\ref{Fig:Combination} we illustrate the constraints arising from the interplay of the three most significant inequalities. In this figure, the allowed area is defined solely by requiring that the OPE predictions exceed the central values of the hadronic moments obtained from lattice QCD inputs.

Finally, because our analysis relies on inequalities that constrain rather than precisely determine the nonperturbative parameters, providing a comprehensive estimate of the associated uncertainties is difficult.  
In principle, such an estimate would also require accounting for $\mathcal{O}(\Lambda_{\text{QCD}}^4)$ contributions and for the scale dependence associated with the choice of $\mu_s$. The former would introduce an additional set of higher-dimensional operators with matrix elements that are currently inaccessible, while the latter source of uncertainty from scale variation is expected to yield only subdominant effects.

\section{Discussion and conclusions}\label{Discussion and conclusions}

Our constraints can be compared with existing estimates of parameters $\hat{\mu}_\pi^2$ and $\hat{\rho}_D^3$, derived from independent methods based on spectroscopic data and the application of the NRCQM.

The kinetic parameter $\hat{\mu}_\pi^2(\Lambda_b)$ can be estimated from spectroscopic data using the HQE for the heavy hadron masses, which leads to the relation~\cite{Bigi:1992su}:
\begin{equation}
    (\overline{M}_D - M_{\Lambda_c}) - (\overline{M}_B - M_{\Lambda_b}) \simeq \left( \frac{1}{2m_c} - \frac{1}{2m_b} \right) (\mu_\pi^2(B) - \mu_\pi^2(\Lambda_b))\,,
    \label{eq:mupi_Lb_determination} 
\end{equation}
with the spin-averaged hadron masses defined as $\overline{M}_H = (M_H+3M_{H^*})/4$. 
This method has been widely used in applications of the HQE to the calculations of heavy baryon lifetimes.
One should note, however, that the relation relies on several assumptions, namely, the equality of $\mu_\pi^2(B)$ and $\mu_\pi^2(D)$ and the equality of the difference $\overline{\Lambda}_{\text{meson}} - \overline{\Lambda}_{\text{baryon}}$ parameters between $b$ and $c$ sectors, as well as the leading-order approximations in the HQE, see e.g. \cite{Gratrex:2023pfn} for details. 

Several analyses of inclusive semileptonic $B \to X_c \ell \nu$ decays~\cite{Alberti:2014yda,Bordone:2021oof,Bernlochner:2022ucr,Finauri:2023kte} have extracted the values of $\mu_\pi^2(B)$ from fits to experimental data. Here, we use the result from Ref.~\cite{Bordone:2021oof}, $\mu_\pi^2(B) = (0.449 \pm 0.042)\,\text{GeV}^2$, obtained in the kinetic scheme at the scale $\mu = 1\,\text{GeV}$. Applying the spectroscopic relation \eqref{eq:mupi_Lb_determination} and adapting the scale, we obtain:
\begin{equation}
\hat{\mu}_\pi^2(\Lambda_b)(\mu=0.75\,\text{GeV}) = (0.43 \pm 0.04)\,\text{GeV}^2\,,
\end{equation} 
consistent with our SVSR constraint shown in Fig.~\ref{Fig:PlotBig}. 

The Darwin parameter, as we already mentioned, can be estimated using the equation of motion for the gluon field strength, which relates it to the matrix elements of spectator dimension-six four-quark operators, up to sub-leading $1/m_b$ corrections \cite{Pirjol:1998ur,Czarnecki:1997sz}. Since the vacuum insertion approximation is unavailable for baryons, the required four-quark matrix elements are estimated using the nonrelativistic constituent quark model with spectroscopic data from heavy baryons as input \cite{Blok:1991st,Neubert:1996we,Rosner:1996fy,Guberina:1997yx,Uraltsev:1996ta}. More detailed discussions and recent applications in determination of lifetimes of heavy baryons can be found in Refs.~\cite{Gratrex:2022xpm,Gratrex:2023pfn}. 
Using this approach, we extracted the value $\hat{\rho}^3_D(m_b)=0.03\,\text{GeV}^3$ in Ref.~\cite{Gratrex:2023pfn}. For the comparison with the current results from Fig.~\ref{Fig:PlotBig}, we evolve this value to the scale $\mu_s=(m_b m_c)^{1/2}$ and convert it to the kinetic scheme at our preferred cutoff scale $\mu=0.75\,\text{GeV}$, which yields
\begin{equation}
    \hat{\rho}^3_{D,\text{kin}}\simeq 0.07\,\text{GeV}^{\,3}\,.
\end{equation}
This value lies slightly below the SVSR region shown in Fig.~\ref{Fig:PlotBig}.

Since the Darwin term enters the decay width with a negative Wilson coefficient, its larger value favoured by our SVSR region would reduce the total decay width of $\Lambda_b$ and thus increase the predicted
ratio $\tau(\Lambda_b)/\tau(B_d)$ given in Ref.~\cite{Gratrex:2023pfn}, shifting it even closer to the measured value.

To summarise, we employed SVSRs for the semileptonic inclusive decay $\Lambda_b \to X_c e^- \bar{\nu}_e$ to relate the moments of hadronic structure functions, evaluated in the OPE, to corresponding expressions involving the hadronic contributions. Truncating the hadronic sum to include the three lowest-lying $\Lambda_c$ states ($J^P = 1/2^+, 1/2^-, 3/2^-$) and using available lattice QCD form factors near zero-recoil as input, we employed inequalities from the zeroth moments as well as from specifically constructed combinations of higher moments (the second and the third). These inequalities were then used to determine the allowed region in the $(\hat{\mu}^2_\pi, \hat{\rho}^3_D)$ plane, as shown in Fig.~\ref{Fig:PlotBig}. We observe the near-saturation of two of the zeroth moments, particularly $\mathcal{I}_{00}^{VV}$, by the ground state $\Lambda_c^{1/2^+}$ contribution within the derived allowed parameter region, consistent with the findings in Ref.~\cite{Mannel:2015osa}. 
This indicates that, in the relevant region of hadronic parameters, the OPE prediction for this moment is already well approximated by the contribution from the exclusive $\Lambda_b \to \Lambda_c^{1/2^+}$ transition alone. As a result, the corresponding sum-rule constraint becomes sensitive to the precision of the lattice QCD input for this exclusive channel. While this near-saturation does not imply an inconsistency, given current uncertainties and poorly known contributions from higher excited states and the continuum—it shows the need for further scrutiny. Further crosschecks in lattice QCD calculations for the semileptonic $\Lambda_b\to \Lambda_c$ form factors are therefore also highly desirable. Initial steps have also been made towards determining the rates of inclusive heavy meson decay rates from fully controlled lattice calculations \cite{Gambino:2022dvu,Kellermann:2025pzt}.

\section*{Acknowledgments}
This work was supported by the Croatian Science Foundation under the project number IP-2024-05-6159. 
We would like to thank Keri Vos for useful suggestions.

\appendix
\section{Hadronic tensor coefficients}
\label{Hadronic tensor coefficients}
In this appendix we collect the expressions for the invariant hadronic tensor coefficients $h_1$, $h_2$, $h_4$, and $h_5$, contributing to $h_{kk}$ and $h_{00}$ in our analysis, evaluated in the operator product expansion (OPE) up to $\mathcal{O}(\Lambda_{\mathrm{QCD}}^3)$ and to leading order in $\alpha_s$ \cite{Gremm:1996df,Koyrakh:1996fr,Dassinger:2006md,Colangelo:2020vhu}. With the definition
\begin{equation}
    z_0\equiv m_b^2-2m_b q_0+q_0^2-\vert \vec{q}\vert^2-m_c^2\,,
\end{equation}
we have, for $\Lambda_b$ baryon in its rest-frame:
\begin{equation}
\begin{split}
    h_1^{VV}&=\frac{1}{z_0}\bigg[q_0+m_c-m_b-\frac{1}{2}\bigg(\frac{1}{3m_b}+\frac{m_c}{m_b^2}\bigg)\hat{\mu}_\pi^2+\frac{1}{2}\bigg(\frac{1}{3m_b^2}-\frac{m_c}{m_b^3}\bigg)\hat{\rho}_D^3\bigg] \\
    &+\frac{1}{z_0^2}\bigg[\frac{3m_c q_0+3q_0^2+2 \vec{q}^{\,2}-3m_b q_0}{3m_b}\hat{\mu}_\pi^2+\frac{3m_c q_0-2m_b^2-m_b(2m_c+q_0)+\vec{q}^{\,2}}{3m_b^2}\hat{\rho}_D^3\bigg]\\
    &+\frac{1}{z_0^3}\bigg[\frac{4}{3}(q_0+m_c-m_b)\vec{q}^{\,2}\hat{\mu}_\pi^2+\frac{4(m_b-q_0)\vec{q}^{\,2}}{3m_b}\hat{\rho}_D^3\bigg]\\
    &+\frac{1}{z_0^4}\bigg[\frac{8}{3}(q_0-m_b)(m_b-m_c-q_0)\vec{q}^{\,2}\hat{\rho}_D^3\bigg]\,,\label{Eq:app1}
\end{split}
\end{equation}
\begin{equation}
\begin{split}
    h_2^{VV}&=\frac{1}{z_0}\bigg[-2m_b-\frac{5}{3m_b}\hat{\mu}_\pi^2-\frac{1}{3m_b^2}\hat{\rho}_D^3\bigg]+\frac{1}{z_0^2}\bigg[-\frac{14}{3}q_0 \hat{\mu}_\pi^2+\frac{2q_0+4m_c-8m_b}{3m_b}\hat{\rho}_D^3\bigg] \\
    &+\frac{1}{z_0^3}\bigg[-\frac{8m_b \vec{q}^{\,2}}{3}\hat{\mu}_\pi^2-\frac{16}{3}(m_b-q_0)q_0\hat{\rho}_D^3\bigg]+\frac{1}{z_0^4}\bigg[-\frac{16}{3}m_b(m_b-q_0)\vec{q}^{\,2}\hat{\rho}_D^3\bigg]\,,\label{Eq:app2}
\end{split}
\end{equation}
\begin{equation}
   \hspace*{-5.6cm} h_4^{VV}=\frac{1}{z_0^2}\bigg[-\frac{4}{3m_b}\hat{\mu}_\pi^2-\frac{2}{3}\frac{\hat{\rho}_D^3}{m_b^2}\bigg]+\frac{1}{z_0^3}\bigg[\frac{8}{3}\frac{q_0-m_b}{m_b}\hat{\rho}_D^3\bigg]\,, \label{Eq:app4}
\end{equation}
\begin{equation}
\begin{split}
    h_5^{VV}&=\frac{1}{z_0}+\frac{1}{z_0^2}\bigg[\bigg(\frac{4}{3}+\frac{5q_0}{3m_b}\bigg)\hat{\mu}_\pi^2+\frac{q_0}{3m_b^2}\hat{\rho}_D^3\bigg]+\frac{1}{z_0^3}\bigg[\frac{4}{3}\vec{q}^{\,2}\hat{\mu}_\pi^2+\frac{4}{3}(2m_b-q_0-\frac{q_0^2}{m_b}\bigg)\hat{\rho}_D^3\bigg]\\
    &+\frac{1}{z_0^4}\bigg[\frac{8}{3}(m_b-q_0)\vec{q}^{\,2}\hat{\rho}_D^3\bigg]\,,
    \label{Eq:app5}
\end{split}
\end{equation}
The coefficients for the axial-axial currents are obtained as
\begin{equation}
    h_i^{AA}=  h_i^{VV}\big\vert_{m_c\to -m_c}\,.\label{Eq:app6}
\end{equation}

\section{Parametrisations of form factors}\label{App: Form factor parametrizations}
In this Appendix we collect the form factor parametrizations for $\Lambda_b\to \Lambda_c$, $\Lambda_b\to \Lambda_c^{\ast\,1/2^-}$ and $\Lambda_b\to \Lambda_c^{\ast\,3/2^-}$, used in QCD lattice evaluations in~\cite{Detmold:2015aaa, Meinel:2021mdj, Meinel:2021rbm}. We also summarize the zero-recoil consistency relations among the form factors, derived in \cite{Hiller:2021zth}, and the standard completeness relations for the Dirac- and Rarita-Schwinger spinors.

\begin{subsection}{$\Lambda_b \to \Lambda_c$}
    For $\Lambda_b \to \Lambda_c$ transition, we adopt the parametrization of the matrix elements from \cite{Detmold:2015aaa}. For the vector current, we have:
\begin{align}
\langle\Lambda_c(p')\vert\bar{c}\gamma^\alpha b\vert \Lambda_b(p)\rangle &= \bar{u}_{\Lambda_c}(p',s')\bigg[f_0^{(\frac{1}{2})^+}\,(m_{\Lambda_b}-m_{\Lambda_c})\frac{q^\alpha}{q^2}\notag\\
    & + f_+^{(\frac{1}{2}^+)}\frac{m_{\Lambda_b}+m_{\Lambda_c}}{s_+}\Big(p^\alpha+p'^\alpha-(m_{\Lambda_b}^2-m_{\Lambda_c}^2)\frac{q^\alpha}{q^2}\bigg)\notag\\
    &+f_{\perp}^{(\frac{1}{2}^+)}\bigg(\gamma^\alpha-\frac{2m_{\Lambda_c}}{s_+}p^\alpha-\frac{2m_{\Lambda_b}}{s_+}p'^\alpha\bigg)\bigg]u_{\Lambda_b}(p,s) \,,\label{FF1}
\end{align}
while for the axial-vector current:
\begin{align}
    \langle\Lambda_c(p')\vert\bar{c}\gamma^\alpha\gamma_5 b\vert \Lambda_b(p)\rangle &=\bar{u}_{\Lambda_c}(p',s')\gamma_5\bigg[-g_0^{(\frac{1}{2})^+}\,(m_{\Lambda_b}+m_{\Lambda_c})\frac{q^\alpha}{q^2}\notag\\
    & - g_+^{(\frac{1}{2}^+)}\frac{m_{\Lambda_b}-m_{\Lambda_c}}{s_-}\Big(p^\alpha+p'^\alpha-(m_{\Lambda_b}^2-m_{\Lambda_c}^2)\frac{q^\alpha}{q^2}\Big)\notag\\
    &-g_{\perp}^{(\frac{1}{2}^+)}\bigg(\gamma^\alpha+\frac{2m_{\Lambda_c}}{s_-}p^\alpha-\frac{2m_{\Lambda_b}}{s_-}p'^\alpha\bigg)\bigg]u_{\Lambda_b}(p,s)\,, \label{FF2}
\end{align}
where $s_\pm=(m_{\Lambda_b}\pm m_{\Lambda_c})^2-q^2$, and $(p,s)$, $(p',s')$ denote the momenta and spin projections of $\Lambda_b$ and $\Lambda_c$, respectively. The only zero-recoil constraint, corresponding to $w=1$ is 
\begin{equation}
    g_\perp^{(\frac{1}{2}^+)}(q^2_{\text{max}})=g_+^{(\frac{1}{2}^+)}(q^2_{\text{max}})\,.\label{FF3}
\end{equation}
\end{subsection}

\begin{subsection}{$\Lambda_b\to \Lambda_c^{\ast\,1/2^-}$}
For the vector current in $\Lambda_b\to \Lambda_c^{\ast\,1/2^-}\ell\nu$ we have
\begin{align}
   \langle\Lambda_c^{\ast\,1/2^-}(p')\vert\bar{c}\gamma^\alpha b\vert \Lambda_b(p)\rangle &= \bar{u}_{\Lambda_c^{\ast\,1/2^-}}(p',s')\gamma_5 \bigg[f_0^{(1/2^-)}(m_{\Lambda_b} + m_{\Lambda_c^{\ast\,1/2^-}})\frac{q^\alpha}{q^2}\notag\\
  &+f_+^{(1/2^-)}\frac{m_{\Lambda_b} - m_{\Lambda_c^{\ast\,1/2^-}}}{s_-}\bigg(p^\alpha+p^{\alpha'}-(m_{\Lambda_b}^2 - m_{\Lambda_c^{\ast\,1/2^-}}^2)\frac{q^\alpha}{q^2}\bigg)\notag\\
  &+f_\perp^{(1/2^-)}\bigg(\gamma^\alpha +\frac{2 m_{\Lambda_c^{\ast\,1/2^-}}}{s_-}p^\alpha-\frac{2m_{\Lambda_b}}{s_-}p^{'\alpha}\bigg)\bigg]u_{\Lambda_b}(p,s)\,,
  \label{FF4}
\end{align}
where $s_\pm=(m_{\Lambda_b}\pm m_{\Lambda_c^{\ast 1/2^-}})^2-q^2$. The corresponding zero recoil constraint is \cite{Hiller:2021zth, Meinel:2021mdj}
\begin{equation}
    f_\perp^{(1/2^-)}(q^2_{\text{max}})=f_+^{(1/2^-)}(q^2_{\text{max}})\,. 
    \label{FF5}
\end{equation}
For the axial-vector current, we have:
\begin{align}
   \langle\Lambda_c^{\ast\,1/2^-}(p',s')\vert\bar{c}\gamma^\alpha\gamma_5 b\vert \Lambda_b(p,s)\rangle &= \bar{u}_{\Lambda_c^{\ast\,1/2^-}}(p',s') \bigg[-g_0^{1/2^-}(m_{\Lambda_b} - m_{\Lambda_c^{\ast\,1/2^-}})\frac{q^\alpha}{q^2}\notag\\
  &-g_+^{(1/2^-)}\frac{m_{\Lambda_b} + m_{\Lambda_c^{\ast\,1/2^-}}}{s_+}\bigg(p^\alpha+p^{\alpha'}-(m_{\Lambda_b}^2 - m_{\Lambda_c^{\ast\,1/2^-}}^2)\frac{q^\alpha}{q^2}\bigg)\notag\\
  &-g_\perp^{(1/2^-)}\bigg(\gamma^\alpha -\frac{2 m_{\Lambda_c^{\ast\,1/2^-}}}{s_+}p^\alpha-\frac{2m_{\Lambda_b}}{s_+}p^{'\alpha}\bigg)\bigg]u_{\Lambda_b}(p,s)\,.\label{FF6}
\end{align}
\end{subsection}

\begin{subsection}{$\Lambda_b\to \Lambda_c^{\ast\,3/2^-}$}
For the vector current in $\Lambda_b\to \Lambda_c^{\ast\,3/2^-}$ transitions, we have 
\begin{align}
\langle \Lambda_c^{\ast\,(3/2^-)}(p')\vert\bar{c}\gamma_\alpha b\vert \Lambda_b(p)\rangle &=\bar{u}^\mu_{\Lambda_c^{\ast\,(3/2^-)}}(p',s')\bigg[f_0^{(\frac{3}{2}^-)}\frac{m_{\Lambda_c^{\ast\,3/2^-}}}{s_+}\frac{(m_{\Lambda_b}-m_{\Lambda_c^{\ast\,(3/2^-)}})p_\mu q_\alpha}{q^2}\notag\\ 
&+f_+^{(\frac{3}{2}^-)}\frac{m_{\Lambda_c^{\ast\,(3/2^-)}}}{s_-}\frac{(m_{\Lambda_b}+m_{\Lambda_c^{\ast\,3/2^-}})p_\mu\Big(q^2(p_\alpha+p'_\alpha)-(m_{\Lambda_b}^2-m_{\Lambda_c^{\ast\,(3/2^-)}}^2 )q_\alpha\Big)}{q^2s_+}\notag\\
&+f_\perp^{(\frac{3}{2}^-)}\frac{m_{\Lambda_c^{\ast\,(3/2^-)}}}{s_-}\bigg(p_\mu\gamma_\alpha-\frac{2p_\mu(m_{\Lambda_b}p'_\alpha+m_{\Lambda_c^{\ast\,(3/2^-)}}p_\alpha)}{s_+}\bigg)\notag\\
&+f_{\perp'}^{(\frac{3}{2}^-)}\frac{m_{\Lambda_c^{\ast\,(3/2^-)}}}{s_-}\bigg(p_\mu \gamma_\alpha-\frac{2 p_\mu p'_\alpha}{m_{\Lambda_c^{\ast\,(3/2^-)}}}+\frac{2p_\mu(m_{\Lambda_b}p'_\alpha+m_{\Lambda_c^{\ast\,(3/2^-)}} p_\alpha)}{s_+}\notag\\
&+\frac{s_- g_{\mu\alpha}}{m_{\Lambda_c^{\ast\,(3/2^-)}}}\bigg)\bigg]u_{\Lambda_b}(p,s)\,,
\label{FF7}
\end{align}
where $s_\pm=(m_{\Lambda_b}\pm m_{\Lambda_c^{\ast 3/2^-}})^2-q^2$.
The zero-recoil constraints are given as \cite{Hiller:2021zth, Meinel:2021mdj}:
\begin{align}
    f_\perp^{(\frac{3}{2}^-)}(q^2_{\text{max}})&=-f_{\perp'}^{(\frac{3}{2}^-)}(q^2_{\text{max}})\,,\notag\\
     f_+^{(\frac{3}{2}^-)}(q^2_{\text{max}})&=-\frac{2(m_{\Lambda_b}-m_{\Lambda_c^{\ast\,(3/2^-)}})}{m_{\Lambda_b}+m_{\Lambda_c^{\ast\,(3/2^-)}}} f_\perp^{(\frac{3}{2}^-)}(q^2_{\text{max}})\,.
     \label{FF8}
\end{align}
For the axial-vector, we have:
\begin{align}
\langle \Lambda_c^{\ast\,(3/2^-)}(p')\vert\bar{c}\gamma_\alpha\gamma_5 b\vert \Lambda_b(p)\rangle &=\bar{u}^\mu_{\Lambda_c^{\ast\,(3/2^-)}}(p',s')\gamma_5\bigg[-g_0^{(\frac{3}{2}^-)}\frac{m_{\Lambda_c^{\ast\,(3/2^-)}}}{s_-}\frac{(m_{\Lambda_b}+m_{\Lambda_c^{\ast\,(3/2^-)}})p_\mu q_\alpha}{q^2}\notag\\ 
&-g_+^{(\frac{3}{2}^-)}\frac{m_{\Lambda_c^{\ast\,(3/2^-)}}}{s_+}\frac{(m_{\Lambda_b}-m_{\Lambda_c^{\ast\,(3/2^-)}})p_\mu\Big(q^2(p_\alpha+p'_\alpha)-(m_{\Lambda_b}^2-m_{\Lambda_c^{\ast\,(3/2^-)}}^2 )q_\alpha\Big)}{q^2s_-}\notag\\
&-g_\perp^{(\frac{3}{2}^-)}\frac{m_{\Lambda_c^{\ast\,(3/2^-)}}}{s_+}\bigg(p_\mu\gamma_\alpha-\frac{2p_\mu(m_{\Lambda_b}p'_\alpha-m_{\Lambda_c^{\ast\,(3/2^-)}}p_\alpha)}{s_-}\bigg)\notag\\
&-g_{\perp'}^{(\frac{3}{2}^-)}\frac{m_{\Lambda_c^{\ast\,(3/2^-)}}}{s_+}\bigg(p_\mu \gamma_\alpha+\frac{2 p_\mu p'_\alpha}{m_{\Lambda_c^{\ast\,(3/2^-)}}}+\frac{2p_\mu(m_{\Lambda_b}p'_\alpha-m_{\Lambda_c^{\ast\,(3/2^-)}} p_\alpha)}{s_-}\notag\\
&-\frac{s_+ g_{\mu\alpha}}{m_{\Lambda_c^{\ast\,(3/2^-)}}}\bigg)\bigg]u_{\Lambda_b}(p,s)\,,
\label{FF9}
\end{align}
with the zero-recoil constraints \cite{Hiller:2021zth, Meinel:2021mdj}:
\begin{align}
    g_{+}^{(\frac{3}{2}^-)}(q^2_{\text{max}})&=g_{\perp}^{(\frac{3}{2}^-)}(q^2_{\text{max}})-g_{\perp'}^{(\frac{3}{2}^-)}(q^2_{\text{max}})\,,\notag\\
    g_{0}^{(\frac{3}{2}^-)}(q^2_{\text{max}})&=0\,. \label{FF10}
\end{align}

The standard completeness relations for the Dirac- and Rarita-Schwinger spinor $\bar{u}^\mu_{\Lambda_c^{\ast\,(3/2^-)}}$ (for the spin-$3/2$ baryon) are:
\begin{align}
    \sum_s u(s)\bar{u}(s)&=\slashed{p}+m \\
    \sum_s u_\mu (s) \bar{u}_\nu(s)&=-(\slashed{p}+m)\bigg(g_{\mu\nu}-\frac{1}{3}\gamma_\mu\gamma_\nu-\frac{2}{3m^2}p_\mu p_\nu-\frac{1}{3m}(\gamma_\mu p_\nu-\gamma_\nu p_\mu)\bigg)\,. \label{FF11}
\end{align}
\end{subsection}

\bibliographystyle{JHEP.bst}
\bibliography{References.bib}

\end{document}